\documentclass[a4paper]{spie}  
\usepackage[]{graphicx}
\usepackage{amssymb,amsmath,psfrag,url}
\usepackage{verbatim,enumitem}

\newcommand{\Bolivarallee}{Boliva\hspace{-0.1mm}r\hspace{0.15mm}a\hspace{-0.1mm}llee}
\newcommand{\Takustrasse}{Taku\hspace{0.25mm}s\hspace{-0.1mm}tra{\ss}e}
\newcommand{\JCMsuite}{{\it JCMsuite\ }}

\title{Advanced finite-element methods for design and analysis of nanooptical structures: Applications}

\author{
Sven~Burger,\supit{\,ab}
Lin~Zschiedrich,\supit{\,a}
Jan~Pomplun,\supit{\,a}
Mark~Blome,\supit{\,b}
Frank~Schmidt\supit{\,ab}
\skiplinehalf
\supit{a}
JCMwave GmbH,
\Bolivarallee~22, 
D\,--\,14\,050 Berlin,
Germany
\smallskip\\
\supit{b}
Zuse Institute Berlin\,(ZIB),
\Takustrasse~7,
D\,--\,14\,195 Berlin,
Germany
}
\authorinfo{
Corresponding author: S. Burger, 
URL: http://www.zib.de, http://www.jcmwave.com,
Email: burger@zib.de.
}

\begin{document}
\maketitle
\noindent
This paper will be published in Proc.~SPIE Vol. {\bf 8642}
(2013) 864205, (DOI: 10.1117/12.2001094), 
and is made available 
as an electronic preprint with permission of SPIE. 
One print or electronic copy may be made for personal use only. 
Systematic or multiple reproduction, distribution to multiple 
locations via electronic or other means, duplication of any 
material in this paper for a fee or for commercial purposes, 
or modification of the content of the paper are prohibited.
Please see original paper for images at higher resolution. 


\begin{abstract}
An overview on recent applications of the finite-element method Maxwell-solver 
\JCMsuite
to simulation tasks in nanooptics is given. 
Numerical achievements in the fields of 
optical metamaterials, 
plasmonics, 
photonic crystal fibers, 
light emitting devices, 
solar cells, 
optical lithography, 
optical metrology, 
integrated optics, 
and photonic crystals 
are summarized.
\end{abstract}




\begin{figure}[b]
\begin{center}
  \includegraphics[width=1.0\textwidth]{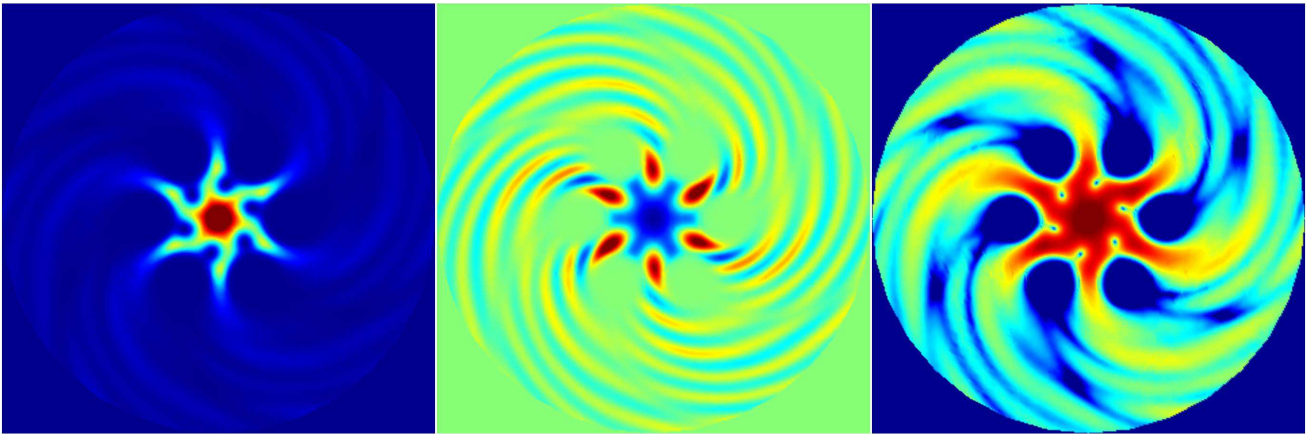}
\caption{
Light intensity distribution of a leaky mode in a twisted photonic crystal fiber (pseudo-color representation). 
From left to right: intensity (linear color-scale), real part of one of the radial electric-field 
vector components (linear color-scale), intensity of the same component (logarithmic color scale).
}
\label{fig_pcf}
\end{center}
\end{figure}

\section{Introduction}
Optical elements with nanometer dimensions are of great importance in many technological 
and scientific research fields. 
Examples are semiconductor device manufacturing (e.g., optical nanolithography), 
new light sources (e.g., VCSELs), diffractive optical elements (DOEs), 
photovoltaics (e.g., thin-film solar cells), sensing (e.g., plasmonic bio-sensors), 
optical communication systems (e.g., integrated optics). 
The functionalities of nanooptical elements critically depend on geometrical and material 
properties of the experimental arrangement. 
For understanding and designing properties of materials and devices 
numerical simulations of Maxwell’s equations are very helpful. 
However, rigorous and accurate simulations of such setups can be challenging because: 
\begin{itemize}[leftmargin=*, topsep = 0pt] 
\setlength{\itemsep}{-4pt} 
\item[-] structures and field distributions are defined on multi-scale geometries (e.g., nanometer layers extending over microns), 
\item[-] material properties (e.g., permittivity of metal) lead to high field enhancements or singularities at edges and corners of the objects, 
\item[-] typical regions of interest are 3D and large in scales of cubic wavelengths, 
\item[-] structures often are embedded into inhomogeneous exterior domains (e.g., plasmonic particles embedded into the material stack of a solar cell).
\end{itemize}
For approaching simulation tasks in nanooptics we develop and use finite-element methods~\cite{monk2003finite} (FEM). 
Main features of FEM are the capability of exact geometric modeling (by using unstructured 
meshes) and high accuracy at low computational cost (due to superior convergence properties of 
higher-order finite elements).
The finite element method offers great flexibility to approximate the solution: 
different mesh refinement levels and polynomial ansatz functions of varying degree  can be combined 
to obtain high convergence rates. 
As a result, very demanding problems can be solved on standard personal computers and workstations.
We demonstrate that the FEM solver \JCMsuite equipped with higher-order finite elements, adaptive
meshing techniques and a rigorous implementation of transparent boundary conditions is 
a powerful method for simulating a variety of settings in nanooptics. 
Here we summarize applications of our FEM developments for simulations of 
several nanooptical devices and applications, 
ranging from fundamental research to industrial development topics. 


\begin{figure}[b]
\begin{center}
  \includegraphics[width=0.99\textwidth]{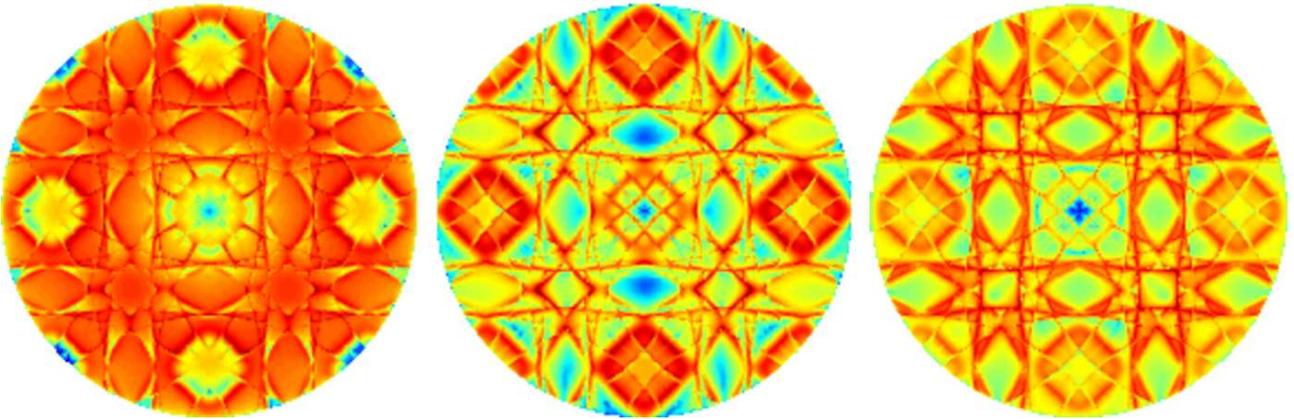}
\caption{
Angular emission spectra of dipoles placed in the emitting layer of an OLED with 
periodically arranged scatterers for improved outcoupling efficiency. 
From left to right: Three different lateral dipole placement positions. 
(Intensity, pseudo-color representation, logarithmic color scale).
}
\label{fig_oled}
\end{center}
\end{figure}


This paper is structured as follows: 
Information on the background of the FEM implementation in \JCMsuite is given in 
Section~\ref{section_background}.
Applications of \JCMsuite to 
optical metamaterials, 
plasmonics, 
photonic crystal fibers, 
light emitting devices, 
solar cells, 
optical lithography, 
optical metrology, 
integrated optics, 
and photonic crystals 
are summarized in
Section~\ref{section_applications}.
The main purpose of this paper is to give an overview on the variety of application fields 
of FEM in nanooptics. 

\section{Background and Methods}
\label{section_background}

The linear Maxwell's equations in frequency domain are an appropriate model to simulate 
optical properties of many technologically relevant 
devices and experiments in fundamental research in the field of nanooptics~\cite{Novotny2006a}.
Three main problem classes can be derived: light scattering problems (including sources like 
plane waves, point sources, or waveguide modes), waveguide mode problems (for geometries which are inviariant in 
one or more space variables), and resonance mode problems~\cite{Pomplun2007pssb,Burger2012springer}. 
Depending on the geometry of the device to be modeled, different coordinate systems are used, 
e.g., 1D, 2D, 3D cartesian coordinates (with periodic, transparent, and/or fixed boundary conditions),  
cylindrically symmetric, or even twisted coordinate systems. 
For computing the electromagnetic near fields in the respective settings we develop and use the 
finite-element (FEM) Maxwell solver \JCMsuite~\cite{Zschiedrich2005a,Burger2008ipnra,Pomplun2007pssb}.
This solver incorporates higher-order edge-elements, self-adaptive meshing, 
and fast solution algorithms for solving time-harmonic Maxwell's equations. 
Also, automatic computation of first- and higher-order parameter derivatives is 
implemented in the software. 
Infinite exterior domains are treated using transparent 
boundary conditions (using an adaptive perfectly matched layer method, 
PML~\cite{Hohage03b,Zschiedrich2006pml,Zschiedrich2006a}).
Further, domain-decomposition (DD) methods are implemented for efficient simulations 
of large 3D computational domains~\cite{Zschiedrich2005b,Schaedle_jcp_2007,Zschiedrich2008al}, 
reduced-basis methods (RBM) have been developed for an online-offline decomposition of 
parameterized simulation setups~\cite{PomplunSCEE2010,Pomplun2010SIAM}, 
and goal-oriented error-estimation is implemented~\cite{zschiedrich2007goaloriented}.


\begin{figure}[b]
\begin{center}
  \includegraphics[width=1.0\textwidth]{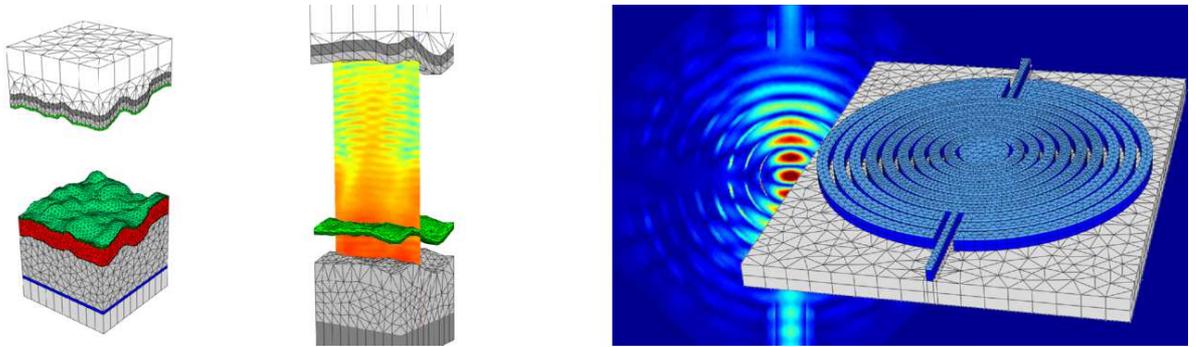}
\caption{
Left: Tetrahedral 3D mesh discretizing the geometry of a thin-film, multi-junction solar cell with 
rough interfaces. Center: Corresponding visualization of a monochromatic light field distribution 
in the solar cell. Right: Prismatoidal 3D mesh discretizing the geometry of a circular-grating resonator 
and corresponding visualization of the excited fundamental mode. 
}
\label{fig_solar}
\end{center}
\end{figure}

\section{Applications}
\label{section_applications}
This section summarizes recent applications of our FEM implementation to simulation tasks in nanooptics. 
Results have been obtained in different academic and industrial research groups and collaborations, 
worldwide. 
Figures~\ref{fig_pcf},~\ref{fig_oled},~\ref{fig_solar} show some exemplary applications: 
Specific properties of a field distribution of a guided mode in a twisted photonic crystal 
fiber~\cite{wong2011strongly,wong2012excitation} are visualized in 
Figure~\ref{fig_pcf}, c.f., Section~\ref{section_pcf}.
Angular emission spectra of light emitting diode with a nano-structured cathode~\cite{zschiedrich2013pw} are displayed in 
Figure~\ref{fig_oled}, c.f., Section~\ref{section_emitters}.
Geometry discretizations and field distributions from solar cell optimization and for an integrated optical 
resonator are shown in 
Figure~\ref{fig_solar}: c.f., Section~\ref{section_solar} for solar cells~\cite{lockau2011rigorous},
c.f., Section~\ref{section_integrated} for Silicon optics simulations~\cite{Burger2010pw2}.

\subsection{Optical metamaterials}
\label{section_metamaterials}
Optical metamaterials are nano-structured materials which can exhibit non-intuitive optical properties, like, e.g., 
a negative refractive index~\cite{Cai2009metamaterials}. 
In this context, the \JCMsuite FEM solver is used to investigate metamaterial building blocks like split-ring 
resonators with resonances at visible frequencies~\cite{Enkrich2005a,Burger2005srr},
magnetic metamaterial properties\cite{Dolling2006a,Linden2006a,Decker2009a,tsakmakidis2010negative},
refractive index properties~\cite{Dolling2007a},
specific resonance properties~\cite{Gansel2010oe,Zhao2011ACS,wong2011strongly}, 
and other effects.

\subsection{Plasmonics}
Plasmonics, or {\it nanoplasmonics}, is the general field of optical phenomena 
related to the electromagnetic response of metals~\cite{Novotny2006a}. 
This includes typical optical metamaterial phenomena, as summarized in the previous Section~\ref{section_metamaterials}. 
However, in the field of {\it plasmonics} typically resonances near metal 
surfaces are the main focus of investigation. 
The sub-wavelength localization of these resonances gives rise to new physical effects and applications. 
Examples are nanoscale lasers and optical sensing at sub-wavelength resolution. 
In this context, \JCMsuite is used to investigate new physical 
effects~\cite{Kalkbrenner2005a,lee2008polarization,Burger2009a,Zhao2011ACS,lee2012excitation}, 
to design plasmonic devices~\cite{unger2009sensitivity,klein2009electron,Lockau2009a,Lockau2009plasmon,Burger2010pw1,burger2010fem,mohammadi2011manipulating,benkenstein2011effects,paetzold2011design,paetzold2011plasmonic,fischer2011plasmon,husakou2011polarization,Paetzold2012,Richter2012PRB,kewes2013design,Yan2012CP} and to 
test theoretical approaches~\cite{Hoffmann2009spie,husakou2011theory,Babicheva2012JOSAB,hiremath2012numerical}.

\subsection{Photonic crystal fibers}
\label{section_pcf}
Photonic crystal fibers (PCF)~\cite{Russell2003a}, or more generally microstructured fibers, are a class of optical fibers 
with specific guiding properties which can be engineered by defining the fiber cross section geometry and the used optical materials. 
This enables a variety of scientific and industrial fields, e.g., frequency-comb-generation, supercontinuum-generation, 
guidance of ultrashort pulses, advanced fiber lasers, and others. 
In this field, \JCMsuite is used to investigate new physical effects in 
PCFs~\cite{Pearce2007oe,lee2008polarization,tyagi2008optical,travers2011ultrafast,wong2011strongly,wong2012excitation}, 
to design PCF for specific functionalities~\cite{Holzloehner2006a,couny2007generation,PoZsKlx07,laegsgaard2008dispersive,Bethge2009jlt,im2009guiding,nold2010pressure,weirich2010liquid,wang2011low,jones2011mid,laurila2011spatial,laurila2011modal}, 
and for further applications. 

\subsection{Light emitting devices}
\label{section_emitters}
Laser diodes and light emitting diodes rely on light emission in the p-n junction of a semiconductor diode, excited by an electric current. 
Applications range from miniaturized light sources to energy-efficient lighting. 
In this field, \JCMsuite is used to investigate and design optical properties of vertical cavity surface emitting lasers 
(VCSEL)~\cite{Rozova2012spie}, 
light emitting diodes (LED, OLED)~\cite{Zschiedrich2012SOLED,zschiedrich2013pw}, 
edge emitters~\cite{Pomplun2010pssb,wenzel2011theoretical,wenzel2013basic}, 
plasmon lasers~\cite{Burger2010pw1,burger2010fem}, 
and other concepts~\cite{Karl2009OE,grossmann2010low,grossmann2010high,grossmann2011strongly,karl2010reversed,jones2011mid,beck2011pmma}. 
In the case of high-power devices, analysis should also include thermo-optical effects~\cite{Rozova2012spie,Pomplun2012thermo,wenzel2013basic}.

\subsection{Solar cells}
\label{section_solar}
Solar cells can convert light to electrical energy. For large-scale electrical power generation, 
thin-film solar cells are advantageous. Microstructures in the different layers of these devices are used
to increase light conversion efficiency. 
Different concepts for so-called {\it light-trapping} rely on regular or rough, metal or dielectric 
nanostructures. 
In this field, \JCMsuite is used to design solar cells for increased conversion efficiency, e.g., by optimizing light trapping 
effects~\cite{Lockau2009plasmon,lockau2011rigorous,benkenstein2011effects,paetzold2011design,paetzold2011plasmonic,Paetzold2012,blome2012back,becker2012large,lockau2013nanophotonic,hammerschmidt20123d}. 

\subsection{Optical lithography}
Photolithography (typically at deep ultraviolet (DUV) and extreme ultraviolet (EUV) wavelengths)
is used for fabrication of patterns on a nanometer scale, with applications especially 
in microelectronics. The field of numerical simulations in this engineering- and research-area is termed 
{\it Computational lithography}~\cite{Lai2012aot}. 
Numerically optimized resolution enhancement techniques (RET), optical proximity correction (OPC), 
and source mask optimization (SMO) help to push the resolution limits of nanofabrication further towards smaller structures.
The technological framework of this field translates to challenging requirements on numerical accuracy and computation time.
In this field, \JCMsuite is used in various industrial 
collaborations~\cite{Burger2005bacus,Burger2006c,Burger2007bacus,Tezuka2007spie,Burger2008bacus,Pomplun2010bacus,Burger2011eom1_,Tyminski2012al}.

\subsection{Optical metrology}
In optical metrology of nanostructures accurate simulation of light propagation is an essential component~\cite{Pang2012aot}. 
A challenge consists in reducing computation times for simulation results matching predefined accuracy requirements such that 
the inverse problems arizing in metrological measurements can be solved {\it online}. 
This is especially important when real-world structures of complex geometry are considered, as it is 
the case in process control and characterization. 
In this field \JCMsuite is mainly used in projects regarding optical metrology of nanostructures of interest to 
the semiconductor industry~\cite{Scholze2007a,Scholze2008a,potzick2008international,Pomplun2008pmj,quintanilha2009critical,Pomplun2009bacusrbm,Burger2011eom1_,Burger2011pm1,Bodermann2011AIP,Bodermann2012op,Kleemann2011eom3,Kato2012a,zang2011structural}.
The FEM implementation is also used by national metrology institutes (PTB, NIST) for critical dimension metrology 
and other purposes~\cite{potzick2008international,quintanilha2009critical,Bodermann2012op,bodermann2012quantitative}.

\subsection{Integrated optics}
\label{section_integrated}
Integrated optical devices ({\it integrated optical circuits}, {Si-optics} devices) integrate several photonic functions into one element. 
This allows for decreasing footprints and in principle for higher performance of standard optical components, e.g., in optical telecommunications, 
and for new functionalities, e.g., for sensing (so called {\it lab-on-a-chip} devices). 
In this field \JCMsuite is mainly used to investigate devices like high-Q resonators, waveguide couplers, splitters, or add-drop 
filters~\cite{Burger2010pw2,Burger2010pw3,Burger2011pw1,warm2011cross,Petracek2012ICTON}. 

\subsection{Photonic crystals}
Photonic crystals are materials with periodic arrangements of the refractive index. 
The specific (periodic or quasi-periodic) arrangements can lead to special properties like the opening of photonic band-gaps. 
Photonic-crystal fibers (see Section~\ref{section_pcf}) are a sub-class of photonic bandgap materials. 
Apart from applications to PCF, in this field \JCMsuite is used to investigate properties of photonic bandgap material 
and devices composed of photonic crystals~\cite{NeveOz2010jap,Burger2005a,Burger2010pw2,Burger2010pw3,Burger2011pw1,Petracek2012ICTON}.

\section{Conclusion}

Adaptive finite-elements prove to be a versatile method for generating accurate results to 
state-of-the-art simulation challenges in nanooptics. 
We have summarized results on analysis, design and optimization of 
nano-structured materials and devices, ranging from fundamental research topics like metamaterials and 
plasmonics to industrial nanooptic applications like microlithography, photonic crystal fibers
and solar cells. 

\section*{Acknowledgments}

This work is supported by BMBF within project {\sc Mosaic} (FKZ 13N12438), by 
Deutsche Forschungsgemeinschaft within DFG research center {\sc Matheon}, and by the European Union
within EMRP Joint Research Project {\sc Ind\,17 Scatterometry}.\\
\includegraphics[width=.2\textwidth]{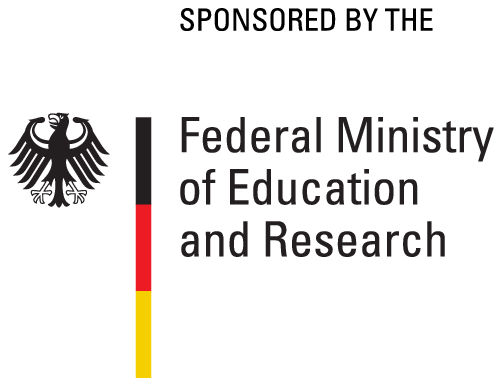}
\includegraphics[width=.5\textwidth]{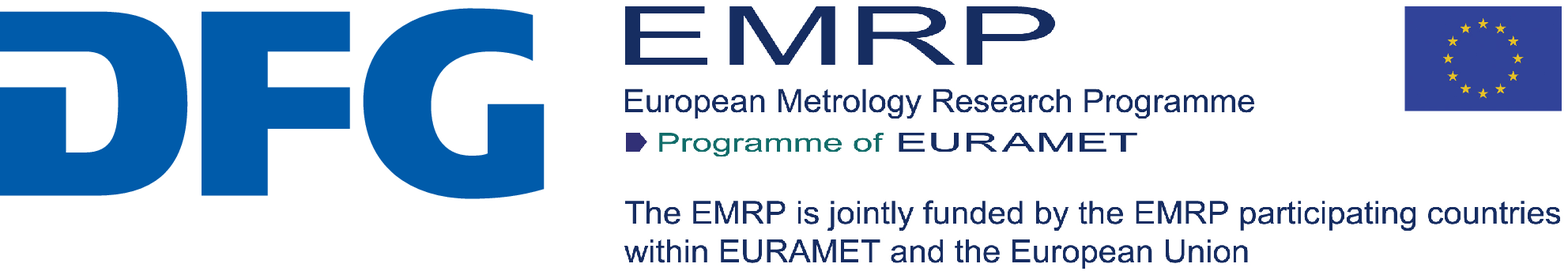}

\bibliography{/home/numerik/bzfburge/texte/biblios/phcbibli,/home/numerik/bzfburge/texte/biblios/my_group,/home/numerik/bzfburge/texte/biblios/lithography,/home/numerik/bzfburge/texte/biblios/jcmwave_third_party}
\bibliographystyle{spiebib}  

\end{document}